\documentclass[12pt]{article}
\usepackage{graphicx}
\usepackage{grffile}
\usepackage{amsmath}

\newcommand\pubnumber{DPF2017-205}
\newcommand\pubdate{}

\def\umn{School of Physics and Astronomy\\
University of California, Irvine, CA 92697, USA}

\def\Title#1{\begin{center} {\Large #1 } \end{center}}
\def\Author#1{\begin{center}{ \sc #1} \end{center}}
\def\Address#1{\begin{center}{ \it #1} \end{center}}

\newcommand\pubblock{\rightline{\begin{tabular}{l} \pubnumber\\
         \pubdate  \end{tabular}}}
\newenvironment{Abstract}{\begin{quotation}  }{\end{quotation}}
\newenvironment{Presented}{\begin{quotation} \begin{center}
             PRESENTED AT\end{center}\bigskip
      \begin{center}\begin{large}}{\end{large}\end{center} \end{quotation}}

\def\beq{\begin{equation}}
\def\eeq#1{\label{#1}\end{equation}}
\def\eeqn{\end{equation}}

\def\beqa{\begin{eqnarray}}
\def\eeqa#1{\label{#1}\end{eqnarray}}
\def\eeqan{\end{eqnarray}}

\let\bar=\overbar

\def\Dslash{\not{\hbox{\kern-4pt $D$}}}
\def\dslash{\not{\hbox{\kern-2pt $\del$}}}

\def\msb{{\bar{\ssstyle M \kern -1pt S}}}

\begin{document}
\begin{titlepage}
\pubblock

\vfill
\Title{Measurement of Neutrino-Electron Elastic Scattering at NOvA Near Detector}
\vfill

\Author{ Jianming Bian \\
(for the NOvA Collaboration)}
\Address{\umn}
\vfill
\begin{Abstract}
NOvA is a long-baseline accelerator-based neutrino oscillation experiment that is optimized for electron-neutrino appearance measurements. It uses the upgraded NuMI beam from Fermilab and consists of a Far Detector in Ash River, Minnesota and a Near Detector at Fermilab. An accurate prediction of the neutrino flux is key to both oscillation and cross-section measurements. The precisely known neutrino-electron elastic scattering cross section provides an in situ constraint on the absolute flux. This talk discusses the status of the measurement of the rate of neutrino-electron elastic scattering in the NOvA Near Detector.


\end{Abstract}
\vfill
\begin{Presented}
DPF 2017\\
The Meeting of the American Physical Society\\
Division of Particles and Fields\\
July 31 - August 4, 2017, Fermilab.\\
C170731\\
\end{Presented}
\vfill
\end{titlepage}
\def\thefootnote{\fnsymbol{footnote}}
\setcounter{footnote}{0}

\section{Introduction}

NOvA (NuMI Off-Axis $\nu_e$ Appearance Experiment)  is a neutrino experiment optimized to observe the oscillation of muon neutrinos to electron-neutrinos \cite{ref:nova}. NOvA uses a 14-kt liquid scintillator Far Detector (FD) in Ash River, Minnesota to detect the oscillated NuMI (Neutrinos at the Main Injector) muon neutrino beam produced 810 km away at Fermilab \cite{ref:numi}. The NOvA baseline is the longest in operation, which maximizes the matter effect and allows a measurement of the neutrino mass ordering. NOvA is equipped with a  0.3-kt functionally identical Near Detector (ND) located at Fermilab to measure unoscillated beam neutrinos and estimate backgrounds at the FD. Both detectors are located 14 mrad off-axis to receive a narrow-band neutrino energy spectrum near the energy of the $\nu_\mu\to\nu_e$ oscillation maximum range ($\sim$2 GeV), enhancing the $\nu_\mu\to\nu_e$ oscillation signal in the FD while reducing neutral current and beam $\nu_e$ backgrounds from high-energy, unoscillated beam neutrinos.

NOvA's detectors consist of plastic (PVC) extrusions filled with liquid scintillator, with wavelength shifting fibers (WLS) connected to avalanche photodiodes (APDs). The dimensions of the detector cells are 6 cm $\times$ 4 cm, with each cell extending the full width or height of the detector, 15.6 m in the FD and 4.1 m in the ND. Extrusions are assembled in alternating layers of vertical and horizontal extrusions plane, so 3-D hit information is available for clustering and particle identification. Each plane (cell width) of the detectors is just 0.15 radiation lengths ($X_0$). This level of granularity helps greatly to separate electrons from $\pi^0$ backgrounds \cite{ref:det1,ref:det2,ref:det3}.

The neutrino flux has a large uncertainty that affects both ND cross-section measurements and FD oscillation analyses in NOvA. Neutrino-electron elastic scattering is a purely leptonic process which can be calculated accurately, so a measurement of $\nu-e$ scattering in the NOvA ND can be used to absolutely constrain the flux. This work will provide a substantial constraint to the flux uncertainty for both ND and FD analyses at NOvA and will demonstrate a flux constraint method for DUNE~\cite{ref:dune}.

In general, $\nu$-e elastic scattering is an elastic two-body collision, and the kinematics can be given by:

\begin{equation}\label{eq:1}
\cos\theta=1-\frac{m_e(1-y)}{E_e},
\end{equation}\label{basic}

where $\theta$ is the angle of the outgoing electron with respect to neutrino beam, $m_e$ is the electron mass, $E_e$ is the electron energy in the final state. $y$ is defined as $y=T_{e}/E_{\nu}$, where $T_{e}$ is the electron kinetic energy and $E_{\nu}$ is the neutrino energy. For neutrino energy at GeV level, Equation~\ref{eq:1} can be approximated as $E_{e}\theta^{2}=2 m_{e}(1-y)$. Since 0 $\leq$ y $\leq$ 1, $E_{e}\theta^{2}<2m_{e}$. Therefore, the signal we are looking for is a single, very forward-going electron with $E_{e}\theta^2$ peaking around zero.

\section{\bf Data sample}\label{xsecgen}

We perform this analysis on an ND data sample consisting of $2.97\times 10^{20}$ Proton-on-Target (POT) exposure. For the neutrino beam, we use FLUKA~\cite{ref:FLUKA} to model hadron production in the NOvA target and use the FLUGG (\cite{ref:FLUGG}) interface to GEANT4~\cite{ref:GEANT4} to simulate the focusing and decay of those hadrons in the NuMI beam. In NOvA detectors, interactions of neutrinos are simulated by the GENIE generator~\cite{ref:GENIE}, and detector responses are simulated by GEANT4. The customized detector simulation chain for NOvA detectors is described in Ref~\cite{Aurisano:2015oxj}.

In GENIE, the cross-section of the $\nu-e$ elastic scattering is calculated at the tree level. To improve the precision of the simulated cross-section, we perform one-loop radiative corrections with modern values of the electroweak couplings to the original GENIE $\nu$-e elastic scattering events. The method is the same as what was used in MINERvA's $\nu$-e elastic scattering paper~\cite{Park:2015eqa}. The size of the signal Monte Carlo simulation (MC) is $6.75\times10^{22}$ POT.

For the background simulation, we use $2.68\times 10^{21}$ POT  inclusive NOvA ND MC sample, composed of $\nu_\mu$ charged current ($\nu_\mu$-CC), $\nu_{e}$ charged current ($\nu_{e}$-CC) and neutral current (NC) events. Meson exchange current events (MEC) \cite{Benhar:2006wy} are added and tuned in this simulation. This sample also includes neutrinos interactions in the rock surrounding the ND.

\section{\bf Event reconstruction} \label{sec:reco}

The $\nu-e$ elastic scattering event reconstruction begins with clustering hits by space-time coincidence to separate beam events from cosmic rays in a trigger window. This procedure can collect together hits from a single neutrino interaction (slice). The slices then serve as the foundation for all later reconstruction stages \cite{ref:reco1}. Next, a modified Hough transform is applied to identify prominent straight-line features in a slice. Then the lines are tuned in an iterative procedure until they converge to the interaction vertex of that slice. Prongs are then reconstructed based on distances from hits to the lines associated with each of the particles that paths emanating from the reconstructed vertex \cite{ref:reco2}-\cite{Niner:2015aya}. We define the shower core based on the prong direction provided by the prong cluster, then collect signal hits in a column around this core. Because the electron deposits energy through ionization in the first few planes then starts a shower, we require the radius to be twice the cell width for the first 8 planes from the start point of the shower and $20$ times the cell width for other planes.

\section{\bf Event selection} \label{sec:evtsel}

Standard NOvA data quality and timing cuts are used to select beam neutrino events under the normal beam and detector conditions. To suppress backgrounds induced by neutrino interactions in the rock (mostly) upstream of the ND, prior to the event identification ($\nu-e$ ID and $e/\pi^0$ ID), we restrict the distance of the primary shower from the detector edges. The containment requirements are more stringent at the front of the detector ($>$ 75 cm) where most of the rock events enter the ND volume. 


To select single-particle events, we require that most of the energy ($>90\%$) in an event is contained by the primary reconstructed shower, no extra energy is deposited around the event vertex (within $\pm$ 8 planes), and that the gap between the start point of the primary shower and the vertex (if a minor, secondary shower is reconstructed) is small ($<$ 20 cm). Our event classifiers have limited performance for low energy backgrounds, so we require that the energy of the primary shower should be greater than 0.5 GeV. In the high energy region, the rate of $\nu_e$-CC is much higher than the $\nu-e$ elastic scattering signal even after the event identification, so a maximum energy of 5.0 GeV is required on the primary shower.

For the event identification, an artificial neural network (ANN) is trained to identify $\nu$-e elastic scattering events ($\nu-e$ ID). The inputs to the ANN are 12 particle likelihood differences between electron and other particle hypotheses ($e-\gamma$, $e-\mu$, etc) for the most energetic shower. Particle likelihoods~\cite{Bian:2015opa} are calculated by comparing the longitudinal and transverse energy deposition in the primary shower to template histograms for simulated $e$, $\gamma$, $\mu$, $\pi^0$, $p$, $n$ and $\pi^{\pm}$.

A $\pi^0$ can be misidentified as an electron by $\nu-e$ ID when only one of the daughter photons is successfully reconstructed, or two daughter photons merge into one shower. To further reject $\pi^0$ backgrounds after the  $\nu-e$ ID selection, a second ANN event selector, $e/\pi^0$ ID, is used. The signature to distinguish an electron from $\pi^0$ backgrounds is the electron's minimum ionization peaks before the multiple scattering (shower) happens, so dE/dx in the first 4 planes are used to form the input of this ANN.

\begin{figure}[h]
\centering
\includegraphics[height=4.5cm]{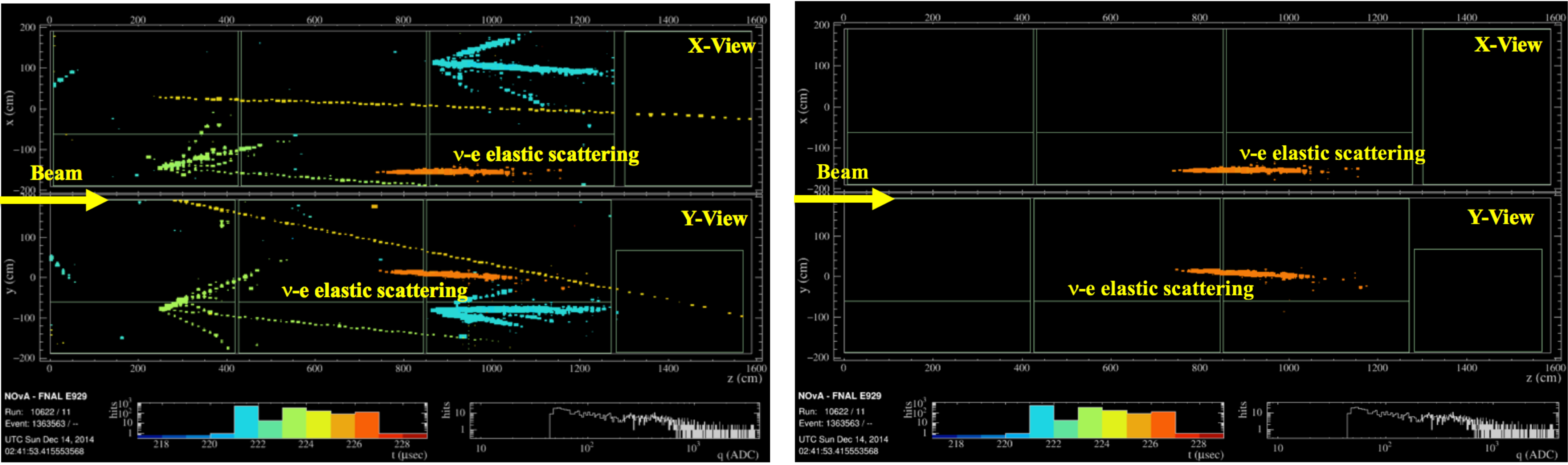}
\caption{Event displays for a simulated $\nu-e$ elastic scattering event in an ND trigger window before (left) and after (right) the event selection.}\label{fig:evtsel}
\end{figure}

Finally, for the most energetic shower in the final state  (electron candidate), the product of energy and the squared angle with respect to the beam ($E_e\theta^2$) must be less than 0.005 GeV$\times$ rad$^2$. Figure~\ref{fig:evtsel} shows event displays for a simulated $\nu-e$ elastic scattering event in an ND trigger window before and after the event selection.

The above selection criteria are chosen to maximize the figure of merit defined as $\frac{S}{\sqrt{(S+B+\delta{B}^{2}})}$, where $S$ and $B$ are the number of signal and background events, and $\delta{B}$ is the systematic uncertainty in the background. We conservatively assume $\delta{B}=0.3B$ according to the $\nu_e$ appearance analyses and the $\nu_e$-CC cross-section measurement at NOvA.

\section{\bf Data Analsys}\label{dataanal}

\begin{figure}[h]
\centering
\includegraphics[height=4.5cm]{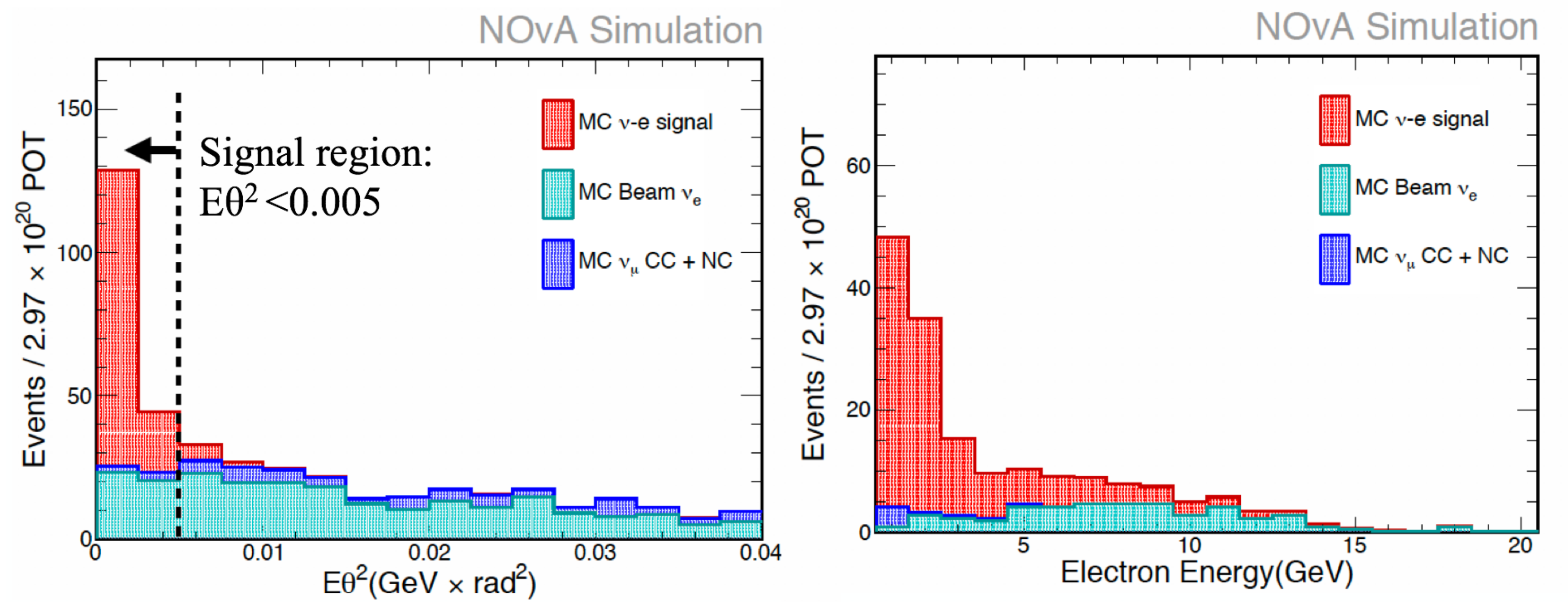}
\caption{$\nu-e$ elastic scattering at NOvA. Left: MC $E_{e}\theta^{2}$ distribution after background correction, Right: MC distribution of electron energy in the $E_{e}\theta^{2}$  signal region after background correction.}\label{fig:nees}
\end{figure}

Figure~\ref{fig:nees} (left) shows the $E_e\theta^2$ distribution of the ND MC events passing all selection criteria except the the $E_e\theta^2$ requirement. $\nu-e$ elastic scattering signal appears as a  peak close to $E_e\theta^2=0$. Due to the result from data is under NOvA collaboration internal review, the data points are not  shown in the figure. In the $E_e\theta^2$ distribution, we define the signal region as $0<E_e\theta^2<0.005$, and the sideband region as  $0.005< E_e\theta^{2} <0.04$. The energy spectrum of the electron candidates in the $E_e\theta^2$ signal region is shown in Figure~\ref{fig:nees} (right).

We use a background subtraction in the $E_e\theta^2$ signal region to extract the yield of $\nu-e$ elastic scattering events. Due to a  Data/MC discrepancy at 10\% level is observed in the sideband region, we correct the yield of the MC background in the signal region ($0<E_e\theta^2<0.005$) by the ratio of signal-subtracted events in the sideband region ($0.005< E_e\theta^{2} <0.04$) in Data and MC. In this background correction, each background component ($\nu_e$-CC, $\nu_\mu$-CC and NC) is scaled by the same factor. Then we take the background-subtracted number of events in the signal region ($0<E_e\theta^2<0.005$) as the yield of $\nu-e$ elastic scattering events.  In the signal region, we expect to see $\sim140$ signal events and $\sim20$ background events in $2.97 \times 10^{20}$ POT ND data, and the statistical error of the signal is $10\%$.

Because the simulation of the $\nu-e$ elastic scattering cross section is precise, the ratio of the $\nu-e$ yields in Data and MC, $N_{\nu-e}(Data)/N_{\nu-e}(MC)$, is majorly determined by the ratio of neutrino fluxes in Data and MC, $\Phi(Data)/\Phi(MC)$.  Since  $N_{\nu-e}(Data)$ and $N_{\nu-e}(MC)$ have been measured in this analysis, the flux in Data can be calculated as:

\begin{equation}\label{eq:b}
\Phi(Data) = \Phi(MC) \times \frac{N_{\nu-e}(Data)}{N_{\nu-e}(MC)}
\end{equation}

\section{\bf Systematic Errors}

\begin{figure}[h]
\centering
\includegraphics[height=4.5cm]{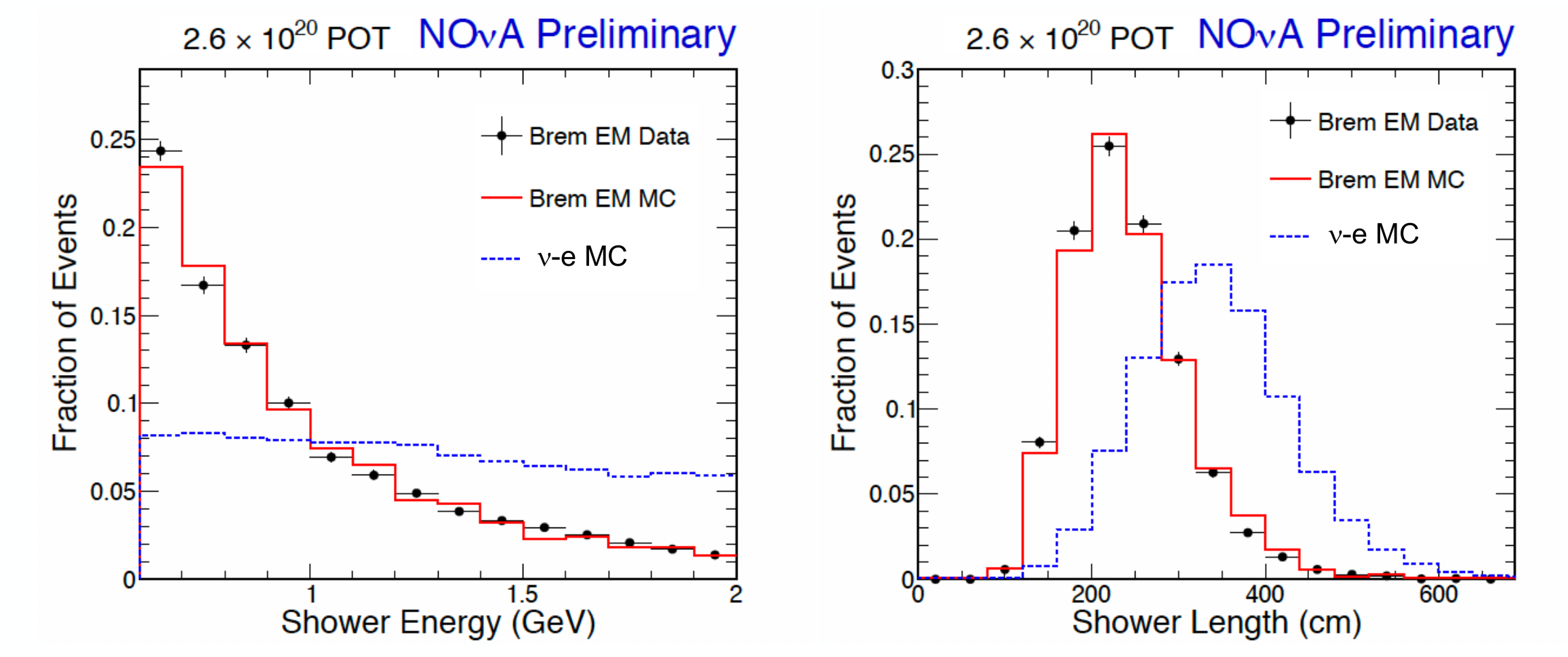}
\caption{Rock muon induced EM showers for efficiency study}\label{fig:mrbrem}
\end{figure}

We use rock muon induced Bremsstrahlung (MR Brem) showers in the ND to estimate the systematic uncertainty in the signal efficiency. Rock muons refer to the muons produced in neutrino interactions in the rock surrounding the ND. Original muons in MR Brem samples are removed using the technique described in Ref~\cite{Sachdev:2013ema} and~\cite{Duyang:2015cvk}. Distributions of energy and length of the MR Brem showers in Data and MC, compared with the $\nu$-e elastic scattering signal MC, are shown in Figure~\ref{fig:mrbrem}. To better mimic selection efficiencies for signal, we re-weight distributions of  $\nu$-e ID, $e/\pi^0$ ID and $E_{e}\theta^{2}$ in MR Brem MC and Data to match corresponding distributions in the $\nu$-e signal MC sample. 

The MR Brem Data/MC differences in the $\nu$-e PID, $e/\pi^0$ PID, and $E_{e} \theta^2$ selection efficiencies are assigned as uncertainties for these selections. While testing one efficiency, cuts on other two variables are applied. The total uncertainty in the signal selection is the quadratic sum of the three terms.  Because we directly check the Data/MC difference in the $E_{e}\theta^{2}$ selection, uncertainties caused by the EM shower angular resolution and the energy calibration are included in this MR Brem shower study. 

Due to inaccuracies in the shower clustering and noise simulation, the single-particle requirement in the event selection could cause an extra uncertainty in the signal efficiency. The MR Brem sample is generated by the muon removal algorithm, which has a different vertex behavior compared with a real neutrino interaction, so we do not use it to validate the efficiency of the single-particle requirement. Instead, we loosen the three cuts on $E_{shower}/E_{tot}$, vertex energy and the gap individually to estimate the uncertainty in each step. The quadratic sum of the three resultant changes is assigned as the systematic error in the single-particle requirement. 

In the background correction, we scale the background interaction modes uniformly according to the sideband data. The systematic error caused by scaling each background component by the same amount is estimated by individually scaling $\nu_e$-CC and NC background components to account for the entire difference between Data and MC in the sideband region. The $\nu_\mu$-CC component only takes a small proportion of the overall background in the signal region, so we do not scale it in this study.

The neutrino interaction cross-section and hadronization uncertainties are determined by reweighting each cross section and hadronization parameter by its uncertainties defined in the GENIE generator. When varying each GENIE parameter, background normalization in the $E_e\theta^2$ signal region is re-calculated by the sideband Data and re-weighted MC.

To estimate the background normalization uncertainty caused by different kinematics in signal and sideband regions, we divide the sideband region ($0.005<E_{e}\theta^2<0.04$) into two sub-regions A and B, and extract background correction factors from them individually. Region A ($0.005<E_{e}\theta^2<0.02$)  is  closer to the  signal region and Region B ($0.02<E_{e}\theta^2<0.04$) is further away. Applying the two scale factors from Region A and Region B to the background MC in the signal region, we assign the larger variation in the signal event counting as the systematic error.

Other systematic errors in the background energy scale, the detector modeling, and the beam intensity simulation are assessed by varying these aspects in the simulation. The analysis is re-performed with these modified MC, and the resultant variation compared to the standard MC is used to quantify each uncertainty. Table~\ref{tab:syst} summarizes systematic uncertainties and their sum in quadrature. The total systematic error in our $\nu-e$ elastic scattering measurement is $6\%$.

\begin{table}
\centering
\includegraphics[height=4.5cm]{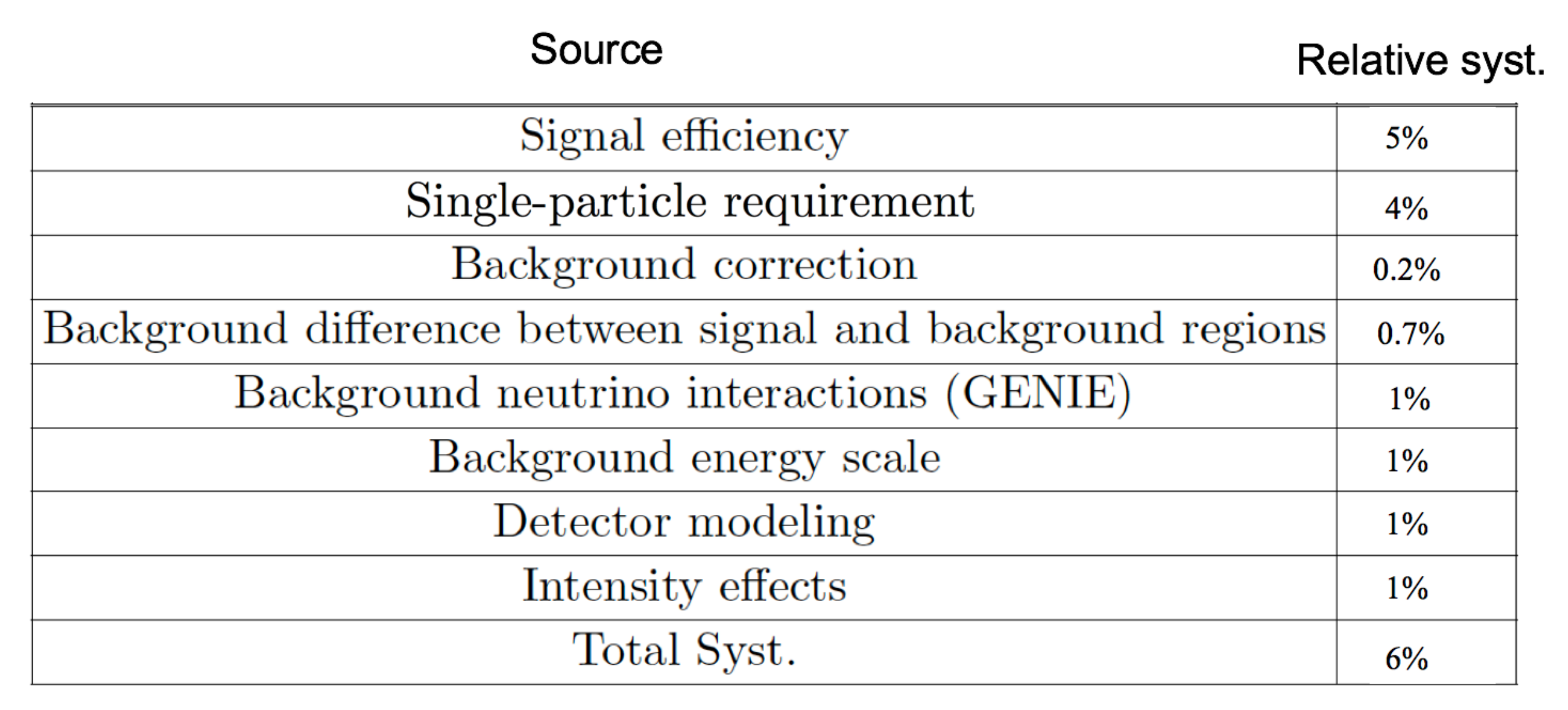}
\caption{Summary of systematic errors}\label{tab:syst}
\end{table}

\section{\bf Summary}
In summary, the $\nu-e$ elastic scattering measurement at NOvA is under way. The ratio of the $\nu-e$ yields in Data and MC, $N_{\nu-e}(Data)/N_{\nu-e}(MC)$, can be used to constrain the NuMI flux. With $2.97\times10^{20}$ POT NOvA ND Data, we expect to see $\sim140$ signal events and $\sim20$ background events. We also estimate the systematic error in $N_{\nu-e}(Data)/N_{\nu-e}(MC)$ to be $\sim6\%$ and the statistical error to be $\sim10\%$.

\end{document}